\documentstyle[preprint,aps,floats,epsfig]{revtex}

\begin{document}

\draft

\preprint{\rightline{ANL-HEP-PR-97-55}}

\title{Topology, fermionic zero modes and flavor singlet correlators in
finite temperature QCD.}

\author{J.~B.~Kogut}
\address{Department of Physics, University of Illinois, 1110 West Green Street,
Urbana, IL 61801, USA}
\author{J.-F.~Laga\"{e} and D.~K.~Sinclair}
\address{HEP Division, Argonne National Laboratory, 9700 South Cass Avenue,
Argonne, IL 60439, USA}

\maketitle

\begin{abstract}
We compute the screening correlators in the $\sigma$ and $\eta^{\prime}$ 
flavor singlet channels in finite temperature QCD with 2 light quark flavors. 
Together with the correlators in the $\vec{\pi}$ and 
$\vec{\delta}$ channels, these are used to discuss several issues
related to symmetry restoration and the nature of the QCD phase
transition. Our calculations span a range of temperature extending from
approximately 125 MeV to 170 MeV and are carried out in the context
of a staggered fermion formulation on a $16^3 \times 8$ lattice.
In addition to the computation at a fixed quark mass ($am_q=0.00625$),
we discuss the issue of the chiral limit. 
After careful consideration of the zero-mode shift lattice artefact, 
we present rather strong (topological) arguments in favor of the 
non-restoration of $U_A(1)$ at $T_c$. 
\end{abstract}


\setcounter{page}{1}
\pagestyle{plain}
\parskip 5pt
\parindent 0.5in

\section{INTRODUCTION}
The lattice approach has been quite successful in describing the general 
aspects of the finite temperature QCD phase transition (for recent reviews 
see \cite{ukawa96,detar,karsch}). 
However the advent of relativistic heavy ion experiments as 
well as purely theoretical motivations calls for even more precise and 
quantitative simulations. Questions such as the determination of the critical 
exponents and the universality class of the phase transition 
(assuming a second order transition, which is favored but not yet proven
\cite{ukawa96}) still have to be answered in detail. Among other things, 
this will 
require simulations at \cite{ks} or close to the chiral limit and may 
necessitate new simulation algorithms. In this paper, we would like to 
delineate a small subset of the issues that one is likely to encounter 
as part of such a program: namely, those questions which are associated 
with the anomalous U(1) axial symmetry. This includes a measurement of 
flavor singlet mesonic correlators together with the extraction of 
flavor singlet susceptibilities and screening lengths (section IV) and lays  
the groundwork for a study of the interplay between topology and the chiral 
phase transition (section V). These topics are closely related to a question 
which has recently attracted much attention in the literature 
\cite{pisarski,shuryak}, namely 
``which chiral symmetry is restored at the finite temperature phase transition 
?''. Attempts at general proofs in the continuum that $U_A(1)$ should be 
restored at $T_c$ \cite{cohen} have been shown to be flawed \cite{lee,evans} 
\footnote{ In this respect, it is also worth mentioning that QED with two 
flavors in 1+1 dimensions provides a counter example to the kind of general 
arguments proposed in \cite{cohen}. There is no spontaneous symmetry breaking 
in D=2, but the effects of the axial anomaly are still manifest for 2 flavors 
as is seen in exact analytical solutions of the massless theory \cite{joos}. }. 
In fact, lattice simulations seem to indicate that this symmetry is only 
restored at higher temperatures \cite{boyd,milc} (although there remain real 
uncertainties concerning the proper method of extrapolation to the chiral 
limit \cite{milc,christ}). In this paper, we identify the zero-mode shift 
phenomenon \cite{vink} as a clear source of difficulties in the chiral limit 
(see section VI). 
We therefore take the position that a rigorous quantitative extrapolation 
to the chiral limit will only be possible once this problem has been solved. 
For the time being, we do two things: first, we work at a fixed but small 
value of the quark mass ($ma=0.00625$ in lattice units) and vary the 
temperature (thereby exploring the direction orthogonal to 
refs.~\cite{milc,christ}). 
Then we use general topological arguments (i.e. the Atiyah-Singer index 
theorem) to decide on the question of restoration or non-restoration of the 
$U_A(1)$ symmetry. We also realize that at present, the investigation of 
topics related to topology is only possible in finite volumes and leave the 
questions related to the extrapolation to the infinite volume limit for 
future studies. 

The measurement of flavor singlet meson correlators and screening lengths 
at finite temperature (section IV) had not been attempted 
previously \footnote{Earlier measurements of 
the ``$\sigma$ meson screening length'' which appeared in the literature 
were in fact representing the $\vec\delta$($\equiv \vec{a_0}$) flavor triplet 
scalar rather than the $\sigma$($\equiv f_0$) flavor singlet scalar, since they only took into account the diagram with connected fermionic lines (and not 
the one with an intermediate pure glue state.)}, but is quite important both 
theoretically and phenomenologically. First, the $\sigma$ (flavor singlet 
scalar meson) is the degree of freedom which becomes light at the transition 
(again assuming a second order phase transition) and therefore drives the 
long distance dynamics together with the pion. 
Second, a determination of the temperature at which 
the $U(1)$ axial symmetry is effectively restored is very interesting because 
it will affect the production rate of $\eta^\prime$ mesons (relative to pions 
for example) in relativistic heavy ion collisions \cite{kapusta,huang}.
Some of the questions considered here have also been investigated through 
different methods: instanton simulations were used in \cite{schafer} and 
Nambu Jona-Lasinio models in \cite{njl}.

In view of the existence of the lattice artefacts mentioned above, we
adopt in this paper a two step strategy to study the restoration of 
symmetries in finite temperature QCD. First, we discuss
the general properties of mesonic correlators in the continuum chiral 
limit (section II). Then we use this as a basis for analyzing 
the implications of our lattice measurements at a non-zero value of the 
quark mass.
The continuum computation allows us in particular to identify the role 
played by topology and fermionic zero-modes. Since $a \rightarrow 0$ and
$m_q \rightarrow 0$ define the ``target'' of symmetry restoration
studies, the general results obtained in this case play an important
role in ``benchmarking'' the actual lattice simulations, which are 
discussed in section III to VI. In section III, we introduce the
parameters of our simulation and discuss some of the techniques used in
the computation. Then in section IV, we present our results for the 
susceptibilities and screening masses as functions of temperature at 
a fixed quark mass ($ma=0.00625$). In section V, we compute the low
lying eigenvalues and eigenvectors of the Dirac operator on our
configuration sample and use them to ``interpret'' the results obtained
for the disconnected correlators in section IV. The issues associated 
with taking the chiral limit are then studied in section VI. In
particular, we show here the importance of ``correcting'' for the 
zero-mode shift lattice artefact. Section VII describes a first attempt
at finding fermionic lattice actions which would have the Atiayh-Singer
index theorem built-in and would therefore allow for a simplified and 
quantitative extrapolation to the chiral limit. Finally, in section VIII
we summarize our results and present our conclusions.

\section{Screening correlators and symmetry restoration}

As is well known, when the $SU(2)_L \times SU(2)_R$ chiral symmetry of QCD
with 2 massless flavors is realized explicitly (rather than being 
spontaneously broken), it implies degeneracies between mesonic correlators. 
In the high temperature phase, we will have for example 
(the signs will be worked out later):
\begin{equation}
|G_{\vec{\pi}}|=|G_{\sigma}| \quad \hbox{and} \quad
|G_{\vec{\delta}}|=|G_{\eta^{\prime}}|
\label{eqn:achdeg}
\end{equation}
where $\sigma$, $\vec{\pi}$, $\vec{\delta}$ and $\eta^{\prime}$ stand 
respectively for the operators $\bar{\psi}\psi$, $\bar{\psi}\gamma_5
\vec{\tau}\psi / \sqrt{2}$, $\bar{\psi} \vec{\tau}\psi / \sqrt{2}$,
$\bar{\psi}\gamma_5\psi$ and $\vec{\tau}$ are the Pauli matrices in 
flavor space with $\psi=( u , d )$. Similarly, if the $U_A(1)$ axial
symmetry were to be effectively restored at high temperatures, we would
have the additional degeneracies:
\begin{equation}
|G_{\vec{\pi}}|=|G_{\vec{\delta}}| \quad \hbox{and} \quad 
|G_{\sigma}|=|G_{\eta^{\prime}}|
\label{eqn:au1deg}
\end{equation}
In other words, all the correlators in the  $\sigma$, $\vec{\pi}$, 
$\vec{\delta}$ and $\eta^{\prime}$ channels become identical if the 
symmetries of both type are restored.

In this section, we will explore in some detail how these degeneracies
come about. This will help us to set the framework for the discussions
that follow. The basic tool that we use is the spectral decomposition of
the quark propagator:
\begin{equation}
S(x,y)=\sum_{\lambda} { \psi_{\lambda}(x) \psi_{\lambda}^{\dagger}(y) 
\over i \lambda + m }
\label{eqn:spdec}
\end{equation}
We will assume a situation where chiral symmetry is restored and 
there is a gap in the eigenvalue spectrum\footnote{ This second 
condition is certainly fulfilled on the relatively small ``boxes'' 
currently considered in lattice simulations. The requirement of a 
finite volume however may not be necessary to its realization (in 
the high temperature phase).}.  
In other words, there will 
be gauge field configurations with exact zero modes (e.g. the
configurations which carry a non-trivial topological charge) or
configurations which don't have any infinitesimally small mode. 
Taking the chiral limit $m \rightarrow 0$ on a finite volume is then 
relatively straightforward (compared to the situation in the broken 
phase where one has to take $V \rightarrow \infty$ first).
When analyzing the flavor singlet correlators ($\sigma$ and
$\eta^{\prime}$) we will have to consider both the connected and
disconnected quark loop contributions. For the flavor triplets 
($\vec{\pi}$ and $\vec{\delta}$), only the connected propagators appear.
In each case, we will have to distinguish between those configurations
which have one zero mode per flavor (and therefore a fermionic
determinant which vanishes like $m^2$ for $N_f=2$) and those which 
have no zero mode.\footnote{Configurations with more than one exact
zero mode per flavor can't contribute in the chiral limit to the average
of mesonic 2-point functions, simply because they come with a fermionic
determinant which vanishes like a higher power of $m^2$ and which can't
be compensated by the maximum of two factors of $1/m$ coming from the 
two quark propagators.}

We start by analysing the connected correlators defined by:
\begin{equation}
C(x,y) \equiv Tr S(x,y)S(y,x)
\label{eqn:C11}
\end{equation}
\begin{equation}
C_{55}(x,y) \equiv Tr \gamma_5 S(x,y) \gamma_5 S(y,x)
\label{eqn:C55}
\end{equation}
Using (\ref{eqn:spdec}), we can write: 
\begin{equation}
S(x,y)= { \psi_0(x) \psi_0^{\dagger}(y) \over m } +
\sum_{\lambda \neq 0} { \psi_{\lambda}(x) \psi_{\lambda}^{\dagger}(y) 
\over i \lambda + m }
\label{eqn:1S1}
\end{equation}
\begin{equation}
\gamma_5 S(x,y) \gamma_5 = { \psi_0(x) \psi_0^{\dagger}(y) \over m } +
\sum_{\lambda \neq 0} { \psi_{\lambda}(x) \psi_{\lambda}^{\dagger}(y) 
\over -i \lambda + m }
\label{eqn:5S5}
\end{equation}
where the first term is present or absent depending on whether there is
or is not an exact zero mode on the configuration considered and we have
used the basic properties of the Dirac operator that the zero modes are
eigenstates of $\gamma_5$ (i.e. either left or right) and that for 
$\lambda \neq 0$: $\psi_{-\lambda}=\gamma_5 \psi_{\lambda}$.
On configurations without zero modes, we will respectively have
for the scalar and pseudoscalar connected correlators:
\begin{equation}
C(x,y) =
Tr \sum_{\lambda \neq 0} { \psi_{\lambda}(x) \psi_{\lambda}^{\dagger}(y) 
\over i \lambda + m }
\sum_{\mu \neq 0} { \psi_{\mu}(y) \psi_{\mu}^{\dagger}(x) 
\over i \mu + m }
\label{eqn:C11nz}
\end{equation}
\begin{equation}
C_{55}(x,y) =
Tr \sum_{\lambda \neq 0} { \psi_{\lambda}(x) \psi_{\lambda}^{\dagger}(y) 
\over -i \lambda + m }
\sum_{\mu \neq 0} { \psi_{\mu}(y) \psi_{\mu}^{\dagger}(x) 
\over i \mu + m }
\label{C55nz}
\end{equation}
and in the chiral limit, we see that on such configurations, the two 
correlators simply differ by a sign: 
\begin{equation}
\lim_{m \rightarrow 0} C(x,y) = - \lim_{m \rightarrow 0} C_{55}(x,y) 
= K(x,y)
\label{Cnz}
\end{equation}
where:
\begin{equation}
K(x,y) = \sum_{\lambda \neq 0} \sum_{\mu \neq 0} 
{ \psi_{\mu}^{\dagger}(x) \psi_{\lambda}(x) \psi_{\lambda}^{\dagger}(y) 
 \psi_{\mu}(y) \over -\lambda \mu} 
\label{eqn:K}
\end{equation}
On configurations with zero modes on the other hand, the first terms in
(\ref{eqn:1S1}) and (\ref{eqn:5S5}) will become dominant at small m and we will get:
\begin{equation}
\lim_{m \rightarrow 0} C(x,y) = \lim_{m \rightarrow 0} C_{55}(x,y) 
= L(x,y)
\label{eqn:Cz}
\end{equation}
where:
\begin{equation}
L(x,y) = 
{ \psi_{0}^{\dagger}(x) \psi_{0}(x) \psi_{0}^{\dagger}(y) 
 \psi_{0}(y) \over m^2} 
\label{eqn:L}
\end{equation}
Note that in this case, there is no change of sign between the scalar
and pseudoscalar correlators, so that the two receive identical
contributions on such configurations.

Then, we consider the disconnected contributions which are
defined by:
\begin{equation}
D(x,y) \equiv Tr S(x,x) Tr S(y,y)
\label{eqn:D11}
\end{equation}
\begin{equation}
D_{55}(x,y) \equiv Tr \gamma_5 S(x,x) Tr \gamma_5 S(y,y)
\label{eqn:D55}
\end{equation}
and constructed from
\begin{equation}
Tr S(x,x)= { \psi_0^{\dagger}(x) \psi_0(x) \over m } +
\sum_{\lambda > 0} { \psi_{\lambda}^{\dagger}(x) \psi_{\lambda}(x) 
\over {\lambda}^2 + m^2 } 2m
\label{eqn:Tr1S}
\end{equation}
\begin{equation}
Tr \gamma_5 S(x,x)= { \psi_0^{\dagger}(x) \gamma_5 \psi_0(x) \over m } +
\sum_{\lambda > 0} { \psi_{\lambda}^{\dagger}(x) \gamma_5 \psi_{\lambda}(x) 
\over {\lambda}^2 + m^2 } 2m
\label{eqn:Tr5S}
\end{equation}
On configurations without zero modes, we get:
\begin{equation}
\lim_{m \rightarrow 0} D(x,y) = \lim_{m \rightarrow 0} D_{55}(x,y) 
= 0
\label{eqn:Dnz}
\end{equation}
whereas on configurations with zero-modes, we get:
\begin{equation}
\lim_{m \rightarrow 0} D(x,y) = \lim_{m \rightarrow 0} D_{55}(x,y) 
= L(x,y)
\label{eqn:Dz}
\end{equation}
i.e. again identical contributions for the scalar and pseudoscalar
correlators in the chiral limit.
Now, we put all of the contributions together and construct
the 2 point functions for the $\sigma$, $\vec{\pi}$, $\vec{\delta}$ and 
$\eta^{\prime}$ in the chiral limit. For this purpose, we have to
remember that disconnected contributions (for the flavor singlets) 
come with a relative factor of $-N_f$ with respect to the connected 
contribution (the - sign comes from Fermi statistics and the factor of
$N_f$ from the 2-loop versus 1-loop). We then get:
\begin{equation}
G_{\sigma}(x,y)= {1 \over Z} \left [ \int_0 K(x,y) + \int_1 L(x,y) 
- 2 \int_1 L(x,y) \right ]
\label{eqn:Gs}
\end{equation} 
\begin{equation}
G_{\vec{\pi}}(x,y)= {1 \over Z} \left [- \int_0 K(x,y) + \int_1 L(x,y) 
\right ]
\label{eqn:Gp}
\end{equation} 
\begin{equation}
G_{\vec{\delta}}(x,y)= {1 \over Z} \left [ \int_0 K(x,y) + \int_1 L(x,y) 
\right ]
\label{eqn:Gd}
\end{equation} 
\begin{equation}
G_{\eta^{\prime}}(x,y)= {1 \over Z} \left [- \int_0 K(x,y) + \int_1 L(x,y) 
- 2 \int_1 L(x,y) \right ]
\label{eqn:Gn}
\end{equation} 
where $\int_n$ represents the functional integral restricted to gauge
field configurations which admit exactly n fermionic zero modes per
flavor and $Z= \sum_n \int_n $. As a reminder, $\int_1$ involves a 
fermionic determinant proportional to $m^2$ which cancels the $1/m^2$ 
in $L(x,y)$ and produces a smooth chiral limit.
From the Atiyah-Singer index theorem:
\begin{equation}
n_L - n_R = Q_{top}
\end{equation}
the sector with one zero mode per flavor is identical to the sector of 
topological charge $\pm 1$, while the sector with no zero-mode is a 
subset of the sector with topological charge 0.

From (\ref{eqn:Gs}-\ref{eqn:Gn}), we find that 
\begin{equation}
G_{\vec{\pi}}=-G_{\sigma} \quad \hbox{and} \quad  
G_{\vec{\delta}}=-G_{\eta^{\prime}}
\label{eqn:chdeg}
\end{equation}
which is simply the consequence of our assumption that chiral symmetry
is restored (i.e. that we could take the $m \rightarrow 0$ limit
naively). We also see that the restoration of the $U_A(1)$ symmetry
\begin{equation}
G_{\vec{\pi}}=-G_{\vec{\delta}} \quad \hbox{and} \quad 
G_{\sigma}=-G_{\eta^{\prime}}
\label{eqn:u1deg}
\end{equation}
would be equivalent to the absense of contribution from the sector with one
fermionic zero mode per flavor and hence of any disconnected contributions. 
This observation will play an important role later when we study the role 
of topology and the link with the Atiyah-Singer index theorem.
At this stage, it is worth mentioning that finite temperature QED in one
space dimension with 2 flavors provides an example of a model where the
integrals appearing in (\ref{eqn:Gs}-\ref{eqn:Gn}) 
can be computed exactly \cite{joos}. As is well
known, there is no chiral symmetry breaking in D=2 \cite{coleman}, although in
the case of $QED_{1+1}$, $T=0$ can be interpreted as a critical point 
\cite{smilverba,smilga} which can only be approached from above. 
In this sense, the
entire phase diagram of $QED_{1+1}$ is mapped (qualitatively) onto the
high temperature phase of $QCD_{3+1}$. What is seen from the $QED_{1+1}$
computation is that in this theory, the configurations with topological 
charge 1 give rise to a non-trivial contribution and the $U_A(1)$
symmetry is not restored (although the $SU(2)_L \times SU(2)_R$ chiral
symmetry is obviously not broken). In the following sections, we will 
attempt the computation of the correlators (\ref{eqn:Gs}-\ref{eqn:Gn}) 
in $QCD_{3+1}$ in the context of lattice gauge theory with staggered quarks.

\section{Parameters and technical aspects of the simulation}

\begin{figure}[htb]
\epsfxsize=4in
\centerline{\epsffile{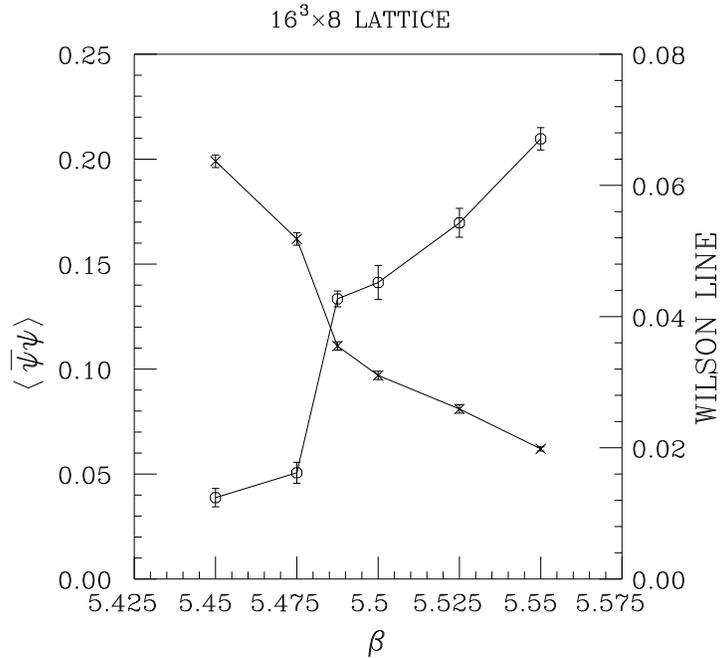}}
\caption{$\langle\bar{\psi}\psi\rangle$ and Wilson line as a function 
of $\beta$ ($N_f=2$,$ma=0.00625$).\label{fig:wilpbp}}
\end{figure}

Our $N_f=2$ simulations were carried out on a lattice of size $16^3 \times 8$ 
and with staggered quarks of mass $ma=0.00625$. The $\beta$ values which were 
studied are $\beta=5.45$, 5.475, 5.4875, 5.5, 5.525 and 5.55 . All of our
configurations were ``borrowed'' from the HTMCGC collaboration \cite{htmcgc8}
except for those at $\beta=5.4875$ which we generated in
order to improve the resolution in the crossover region. Using the
formulae given in \cite{beta}, we identify the above values of $\beta$ with
the temperatures: $T \approx 125$, 135, 140, 145, 157 and 170 MeV. The diagram
for the chiral condensate and the Wilson line (containing the data from
\cite{htmcgc8} together with the new point at $\beta=5.4875$) is presented in
fig. \ref{fig:wilpbp}. The crossover at this value of the quark mass is 
now placed between $\beta=5.475$ and $\beta=5.4875$.

In addition to these full QCD simulations, we have also carried out a
few quenched computations in order to clarify some technical aspects. 
We used the same lattice size ($16^3 \times 8$) and considered the
$\beta$ values $\beta=5.8$, 5.9, 6.0, 6.1 and 6.2 . The phase transition
which is expected to be first order in this case occurs
around $\beta=6.0$ .

In all of the computations of the mesonic correlators described below,
the operators representing the $\vec{\pi}$, $\sigma$, $\vec{\delta}$ and
$\eta^{\prime}$ mesons are respectively taken to be 
$\gamma_5 \otimes \xi_5$, $ I \otimes I $, $ I \otimes \xi_5$ and 
$\gamma_5 \otimes I$ (in standard ``spin $\otimes$ flavor'' staggered 
fermion notation). The first two of these operators are local and 
the last two 4-link operators. The connected and disconnected pieces 
of the correlators were computed using a $U(1)$ noisy estimator
following the techniques used by Kilcup et al. \cite{kilcup,venka} in zero
temperature QCD. 

In order to help in the interpretation of our results, we also computed
the low lying spectrum of the Dirac operator (in practice the lowest 8
eigenvalues and associated eigenvectors) on each of our configurations.
This was done using a conjugate gradient algorithm in the form
investigated by Kalkreuter and Simma in ref \cite{kalk}
.\footnote{ It is also worth mentioning that some improvements over this
technique \cite{bunk} as well as other techniques \cite{sorensen} have recently been
used successfully \cite{venka,negele} to compute a larger number of eigenvectors (currently
up to 128).} 

We have analysed 160, 240, 140, 160, 160 and 80 configurations at 
$\beta=5.45$, 5.475, 5.4875, 5.5, 5.525 and 5.55 respectively. 
These configurations are spaced by 5 units of molecular dynamics time 
(except for those at $\beta=5.55$ which are spaced by 10 units) 
\cite{htmcgc8}.
The disconnected correlators were computed using a $U(1)$ noisy source
spread over the entire volume. From 8 to 32 random sources were used per
configuration. 8 were found to be sufficient in general for our purposes
and were used in the bulk of our computations. The correlators were then
measured in the directions x, y and z and averaged over direction. 
The connected correlators were computed using a $U(1)$ source defined on
a given z-slice, or 2 adjacent z-slices for the 4-link operators. In the
later case, a second inversion was carried out after ``transporting'' the 
source across the hypercube as prescribed by the form of the non-local
staggered operator. The whole procedure was repeated on 2 slices
separated by distance 8, or 4 slices separated by distance 4.
Alternatively, we used 1 slice in each of the 3 x, y and z directions 
which gives equivalent or even better results.

\section{Susceptibilities and screening lengths}

\begin{figure}[htb]
\epsfxsize=4in
\centerline{\epsffile{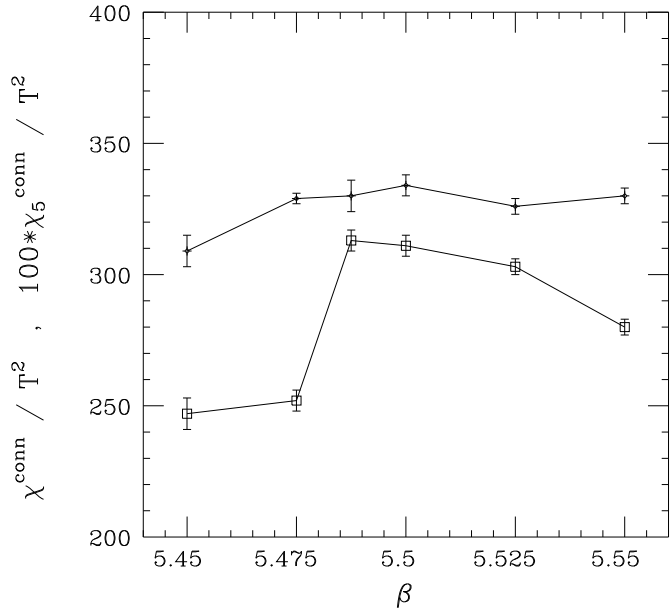}}
\caption{Connected susceptibilities as a function 
of $\beta$ ($N_f=2$,$ma=0.00625$).\label{fig:suscon}}
\end{figure} 

\begin{figure}[htb]
\epsfxsize=4in
\centerline{\epsffile{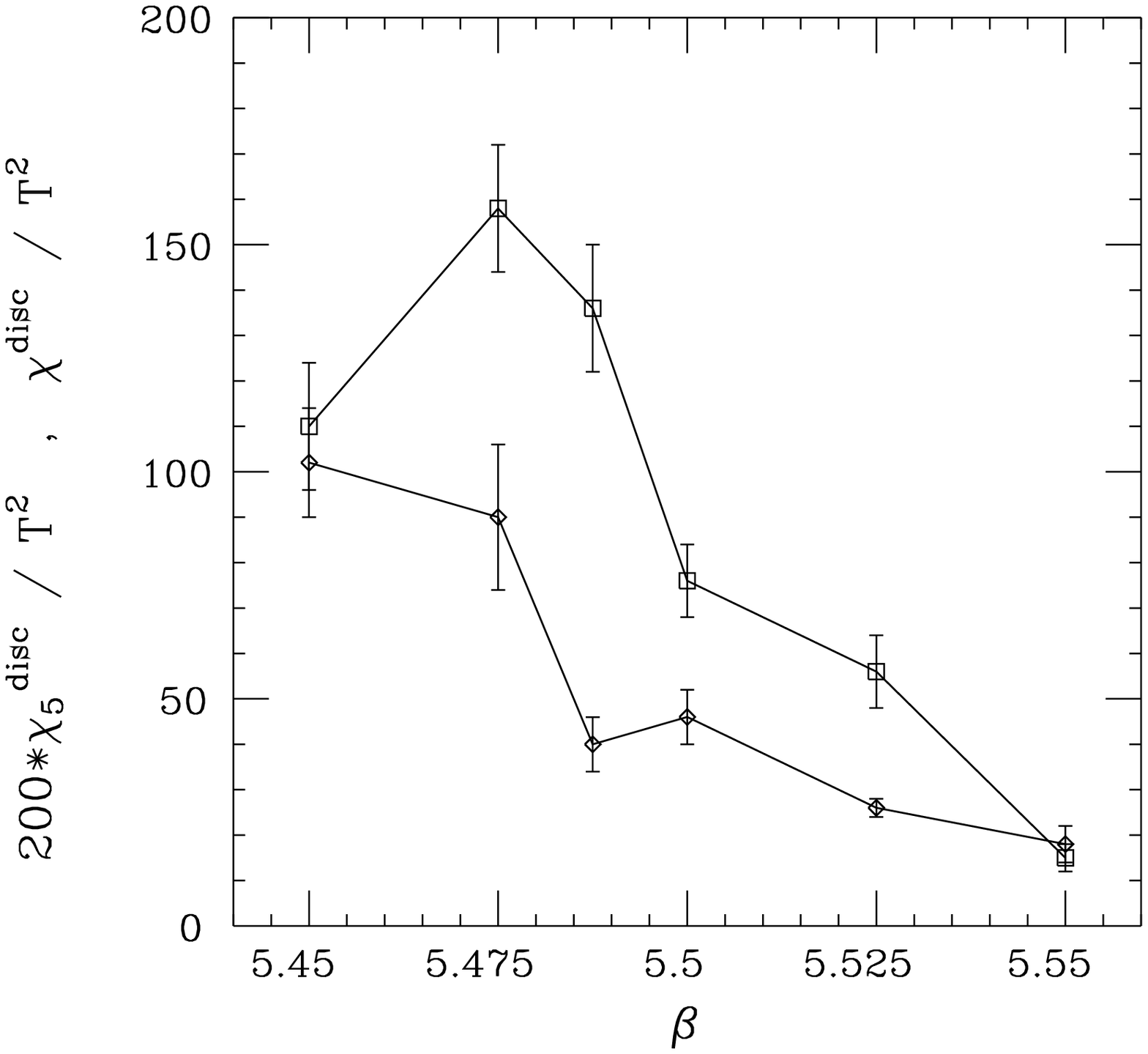}}
\caption{Disconnected susceptibilities as a function 
of $\beta$ ($N_f=2$,$ma=0.00625$).\label{fig:susdis}}
\end{figure} 

To start with, we consider our results for the scalar and pseudoscalar
susceptibilities. We separate the connected (i.e. volume integral of 
equations \ref{eqn:C11} or \ref{eqn:C55}) and disconnected (i.e. volume 
integral of equations \ref{eqn:D11} or \ref{eqn:D55})
contributions which are represented respectively as functions of
$\beta$ in fig.~\ref{fig:suscon} and fig.~\ref{fig:susdis}. 
We will first discuss the scalar
susceptibility. Our results at $N_t=8$ are compatible with those of
earlier works at $N_t=4$ \cite{karsch4}, $N_t=6$ \cite{milc6} and $N_t=12$ \cite{milc12}
.\footnote{See ref. \cite{milc12} for a summary plot of earlier measurements.}
In particular, we find a peak in the disconnected but not in the
connected susceptibility. It is important to note that if this situation
were to persist in the chiral limit it would imply that only the flavor
singlet scalar becomes massless at the transition and not the flavor
triplet scalar. This would imply that the $U(1)$ axial symmetry remains 
broken at $T_c$. We will return to this discussion below. At a more
technical level, we also find some slight differences with earlier
works. For example, in the connected scalar susceptibility, we see a larger
jump at the crossover than had been seen before. This is consistent
with a rather abrupt change of the screening mass of the $\delta$ meson
(see below), although the later effect is not statistically as
significant. In the pseudoscalar channel, the most interesting behavior
is related to the disconnected correlator. This is because of its
association with topology through the integrated anomalous Ward
identity: 
\begin{equation}
m \int d^4 x \bar{\psi} \gamma_5 \psi = Q_{top}
\label{eqn:iward}
\end{equation}
on each gauge field configuration. Therefore, in the continuum, we would have:
\begin{equation}
m^2 \chi_5^{dis} = \chi_{top} \equiv { < Q^2_{top} > \over V }
\label{eqn:xtop}
\end{equation}
As we will see below, this relation doesn't translate exactly to the
lattice. However a remnant can be identified.
We also note that the connected pseudoscalar susceptibility is almost 
constant accross the transition. This is in part a lattice artefact. 
In the continuum, we would expect that this susceptibility (equivalent
to the flavor triplet pseudoscalar susceptibility) picks up a large
contribution in the broken phase from the near masslessness of the pion
in the chiral limit. This does not occur here because the lattice 
$\Gamma_5$ operator used above is not associated with a Goldstone 
pion in the staggered formulation.

\begin{figure}[htb]
\epsfxsize=4in
\centerline{\epsffile{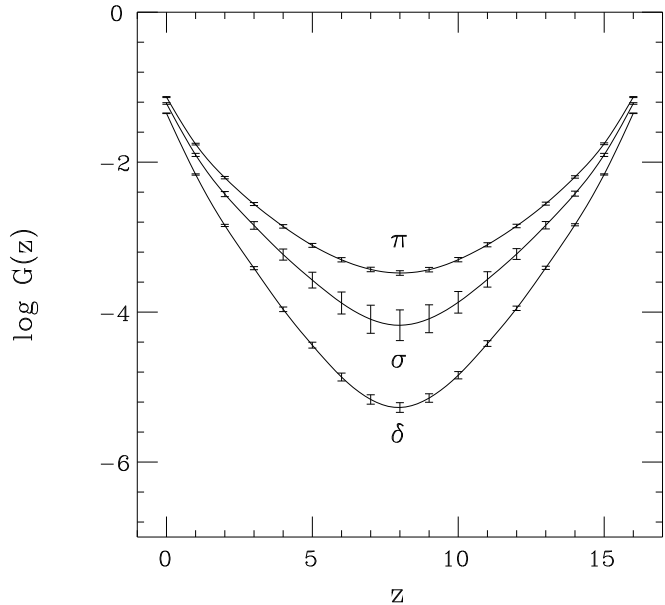}}
\caption{Screening correlators at $\beta=5.4875$ ($N_f=2$,$ma=0.00625$).
\label{fig:sig5p4875}}
\end{figure} 

\begin{figure}[htb]
\epsfxsize=4in
\centerline{\epsffile{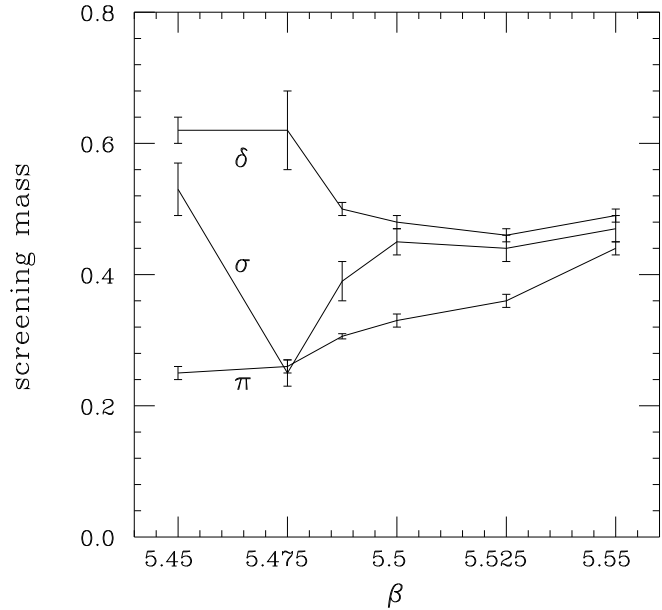}}
\caption{Screening masses as a function 
of $\beta$ ($N_f=2$,$ma=0.00625$).\label{fig:sigfull}}
\end{figure} 

\begin{figure}[htb]
\epsfxsize=4in
\centerline{\epsffile{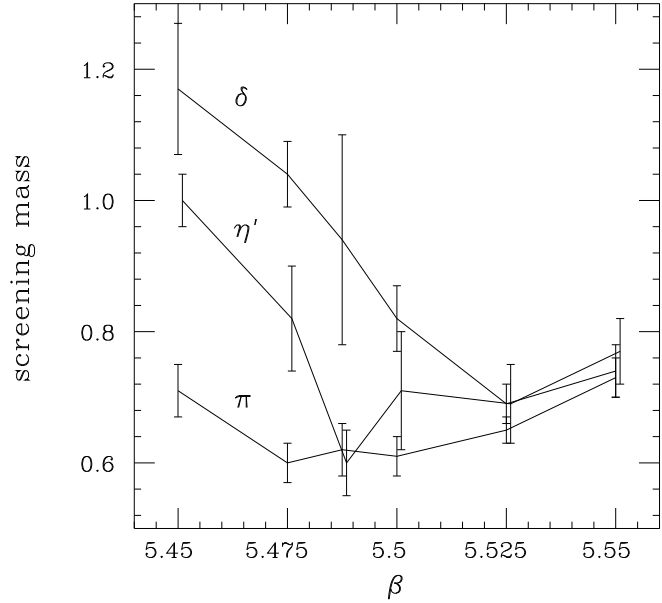}}
\caption{Screening masses in the channels associated with 4-link mesonic 
operators as a function 
of $\beta$ ($N_f=2$,$ma=0.00625$).\label{fig:psfull}}
\end{figure}

We now discuss the measurements of the screening masses. Because of
large flavor symmmetry breaking, it is useful to separate the mesonic 
correlators into two categories: those corresponding to local operators
and those corresponding to four link operators. In the first category,
we have the pion (Goldstone) and the $\sigma$, to which we can add
as well a representative of the $\vec{\delta}$ (connected part of the
$\sigma$). Similarly, in the second category we would have the
$\eta^{\prime}$ and $\vec{\delta}$ as well as a non-Goldstone pion 
(connected part of the $\eta^{\prime}$).
In fig.~\ref{fig:sig5p4875} we show an example of correlators belonging
to the first category at $\beta=5.4875$ (i.e. right above the crossover
induced by the chiral phase transition). The screening masses extracted 
from our fits to
those correlators as well as those measured at other values of $\beta$
are shown in fig. \ref{fig:sigfull}. 
The key feature of this plot is that the $\sigma$ becomes light
close to the transition while the $\vec{\delta}$ remains heavy. This is
what we would expect if the $U_A(1)$ symmetry were not restored at the
chiral phase transition. It is also in agreement with the observation 
made earlier that the peak in the scalar susceptibility originates
in the disconnected part of the correlator (see above). It is worth 
mentioning here that some of the fits leading to fig. \ref{fig:sigfull} 
may have large
systematic errors (only the statistical errors are included in the
figure). This is due in part to the small extent of the lattice 
(and the use of point sources rather than ``smearing improved'' sources)
and, in the case of the $\sigma$, to the additional difficulties
associated with the measurements of disconnected quark loop correlators.
Nevertheless, the qualitative picture emerging from fig. \ref{fig:sigfull} 
appears rather clear: the $\sigma$ becomes lighter close to the transition 
while the $\vec{\delta}$ remains heavy. 
In fact, the trends observed in the early reports on
this work \cite{lagae} have been further confirmed by our recent addition of a 
``data point'' at $\beta=5.4875$ . The main question that remains is
the problem of the extrapolation to the chiral limit. Certainly, 
there are still rather large explicit chiral symmetry breaking effects
in our current data ($ma=0.00625$) as can be seen from the fact that 
above the chiral phase transition ($m_{\sigma}-m_{\vec{\pi}}$) is almost
as large as the $U_A(1)$ symmetry breaking ($m_{\vec{\delta}}-
m_{\vec{\pi}}$). Measurements at lower values of the quark mass would 
therefore be needed to clarify the situation but are beyond the scope 
of this work. In addition,
in section VI, we show that the issue of the chiral limit is rather
subtle and requires a detailed understanding of at least some lattice
artefacts. This implies in particular that a quantitative determination
of the screening masses in the chiral limit will require the use of
improved actions or very large lattices. 
In summary, the data shown in fig. \ref{fig:sigfull} are suggestive of
a situation where $U_A(1)$ is only restored at some $T>T_c$, but not 
sufficient by themselves to prove this fact. We will only achieve that
goal after identifying the topological origin of the difference between
the $\vec{\delta}$ and $\sigma$ propagators for small quark masses in the high 
temperature symmetric phase (see section VI). 
Finally, in fig.~\ref{fig:psfull}, we also present for completeness the 
screening
masses obtained from the correlators involving four-link operators. 
These however are much less informative at the current values of
$\beta$, since flavor symmetry breaking makes all the states heavy in 
this case. Note that fig.~\ref{fig:psfull} is drawn to the same scale as
fig.~\ref{fig:sigfull} but with a mass shift of $0.5$ along the vertical
axis.

\section{Low lying fermionic modes and disconnected correlators}

As was shown in section II, the presence of $U_A(1)$ symmetry breaking 
effects at $m=0$ in the $SU(2)_L \times SU(2)_R$ symmetric phase is 
equivalent to the existence of a non-zero disconnected contribution 
to flavor singlet correlators (compare formulae (\ref{eqn:Gp}) and 
(\ref{eqn:Gd}) for example). In addition,
these contributions are accounted for entirely by exact fermionic 
zero-modes (\ref{eqn:L}). 
An interesting way of studying $U_A(1)$ breaking in the chiral
limit is therefore to compute the low lying eigenmodes of the Dirac
operator. In this section, we compute the lowest 8 (positive)
eigenvalues ($\lambda$) and the associated modes on each configuration 
of our sample at $ma=0.00625$ 
and discuss how close they are to satisfying the Atiyah-Singer index 
theorem and how well they already saturate the disconnected correlators.
In the next section, we will see how the knowledge of these low modes 
can be used to extract information about the chiral limit. In both
cases, the existence of the zero-mode shift lattice artefact \cite{vink} 
requires a detailed and careful analysis.

\begin{figure}[htb]
\epsfxsize=4in
\centerline{\epsffile{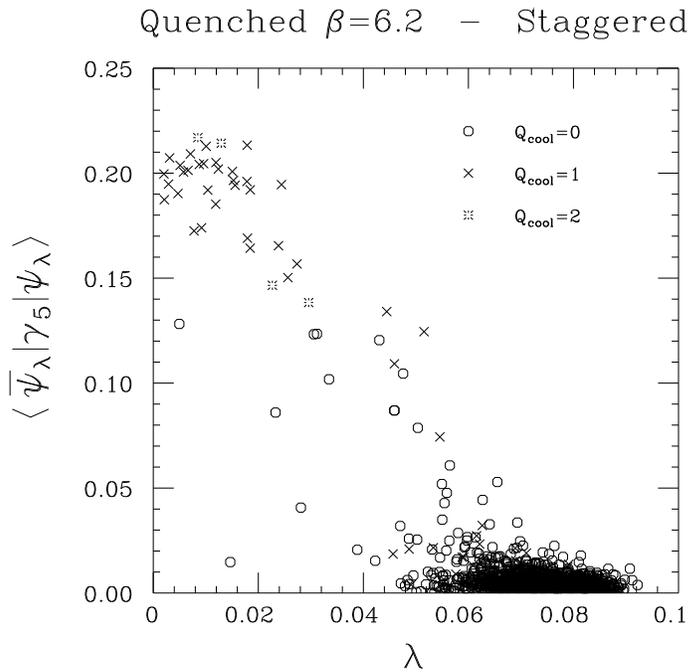}}
\caption{Eigenmode chirality versus $\lambda$ for the lowest 8
eigenvalues on each configurations of our quenched sample at $\beta=6.2$ 
.\label{fig:rl6p2}}
\end{figure} 

\begin{figure}[htb]
\epsfxsize=4in
\centerline{\epsffile{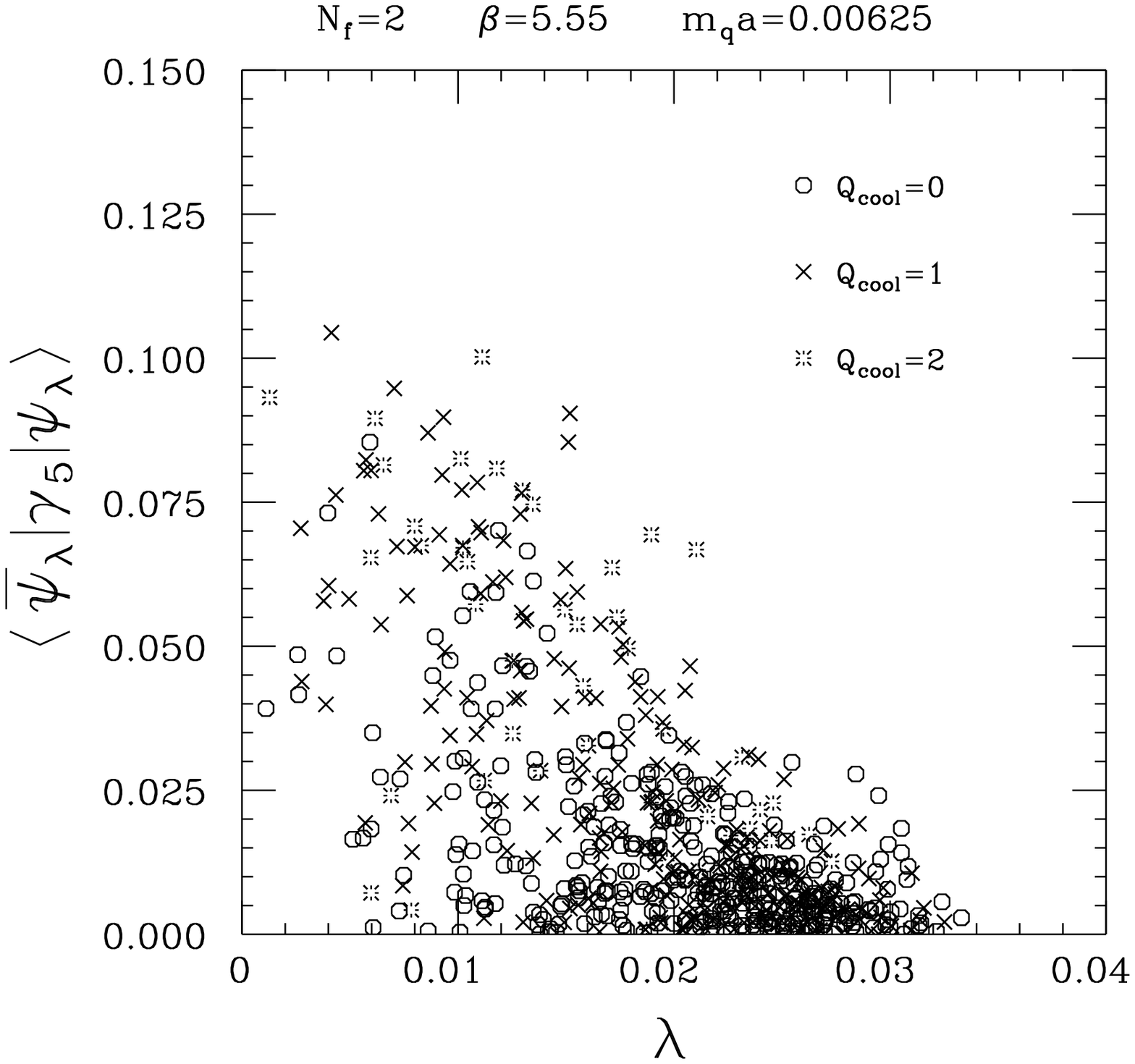}}
\caption{Eigenmode chirality versus $\lambda$ for the lowest 8
eigenvalues on each configurations of our full QCD sample at $\beta=5.55$ 
.\label{fig:rl5p55}}
\end{figure} 

We start by studying the disconnected susceptibilities (integrated
correlators). For this purpose, we make use of the spectral
decomposition of the quark propagator (\ref{eqn:spdec}) and the resulting 
formulae:
\begin{equation}
Q \equiv Tr S = \sum_{\lambda > 0} { 2m \over \lambda^2 + m^2 } 
+ { n_L + n_R \over m }
\label{eqn:Q}
\end{equation}
\begin{equation}
Q_5 \equiv Tr \gamma_5 S = { n_L - n_R \over m }
\label{eqn:Q5}
\end{equation}
where $n_L(n_R)$ are respectively the number of left (right) zero-modes
of the Dirac operator. In terms of these, the scalar and pseudoscalar 
disconnected susceptibilities are defined as:
\begin{equation}
\chi^{dis}= [ <Q^2>-(<Q>)^2 ]/V
\label{eqn:C1d}
\end{equation}
\begin{equation}
\chi_5^{dis}= <Q^2_5>/V
\label{eqn:C5d}
\end{equation}

\begin{figure}[htb]
\epsfxsize=4in
\centerline{\epsffile{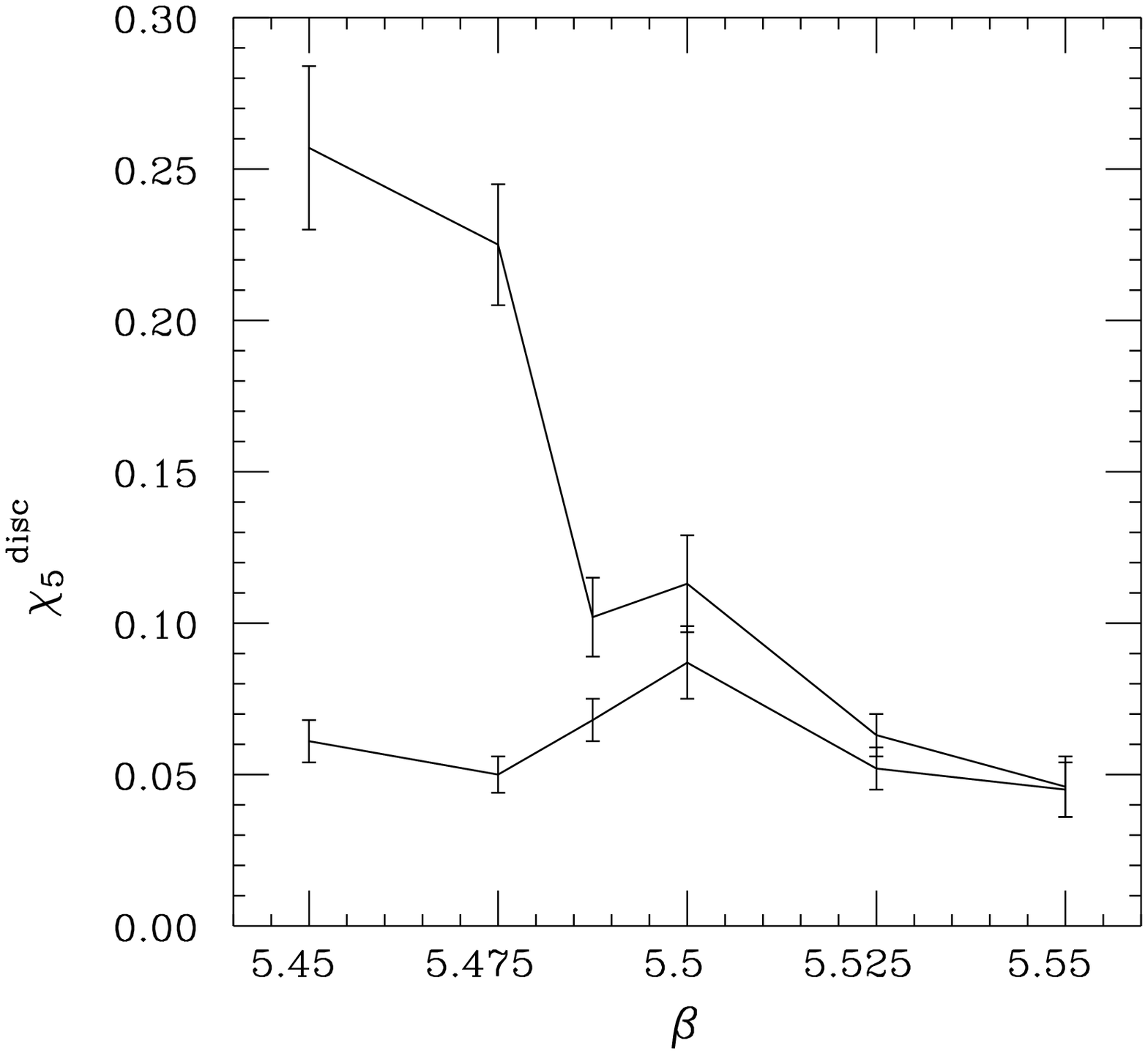}}
\caption{Disconnected pseudoscalar susceptibilities as a function 
of $\beta$. Complete result (top curve) and truncation to the lowest 
8 modes (bottom curve).\label{fig:pszmdis}}
\end{figure} 

\begin{figure}[htb]
\epsfxsize=4in
\centerline{\epsffile{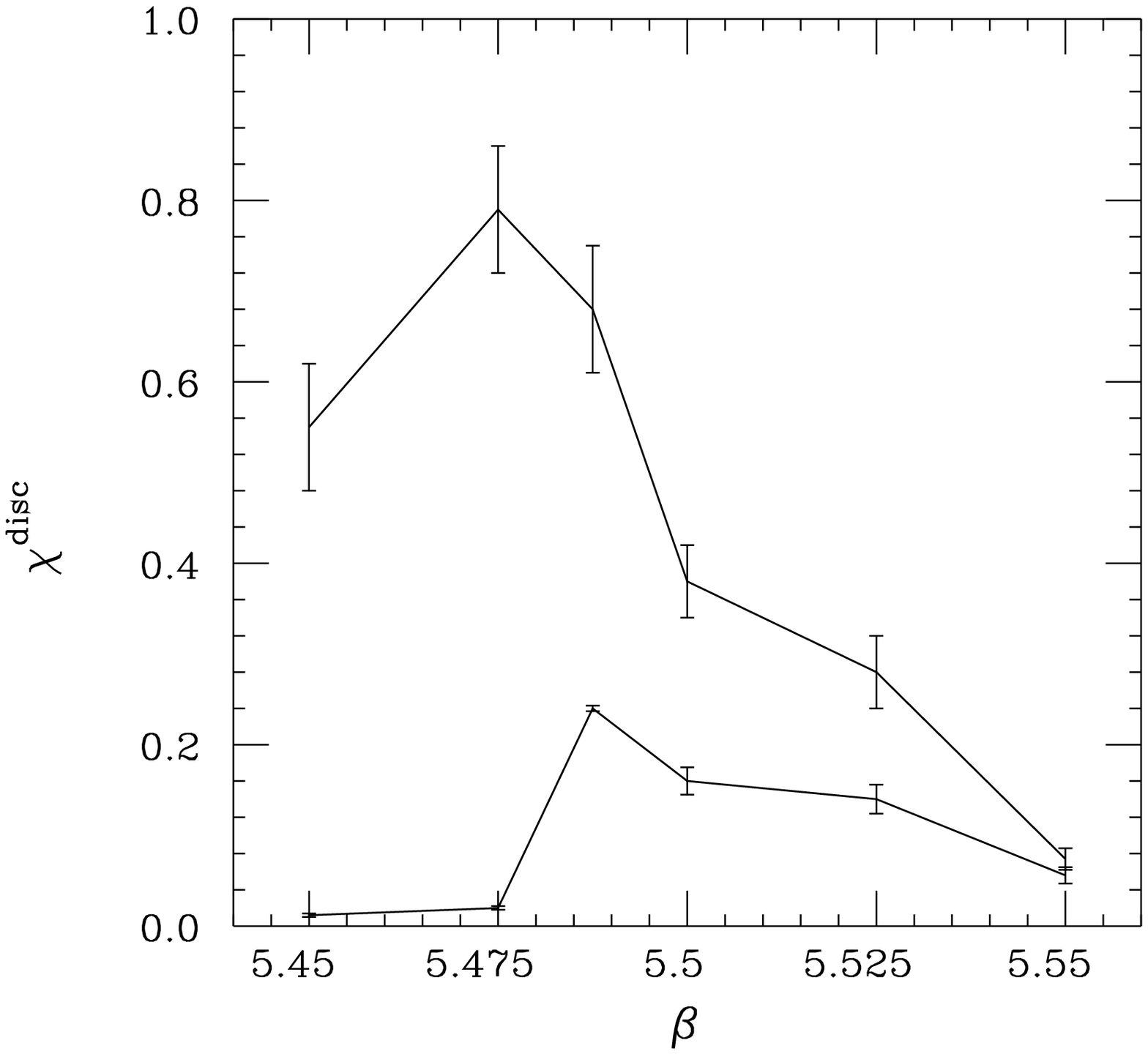}}
\caption{Disconnected scalar susceptibilities as a function 
of $\beta$. Complete result (top curve) and truncation to the lowest 
8 modes (bottom curve).\label{fig:sczmdis}}
\end{figure} 

Following the same reasoning as in section II, it is shown that in the 
continuum, $\chi_{(5)}^{dis}$ is completely saturated in the chiral limit
by contributions from configurations with one zero-mode per flavor. 
On the lattice however this situation is not reproduced exactly. First
there are no exact zero-modes (except on a subspace of measure 0). Then: 
\begin{equation}
Q^{latt}=\sum_{\lambda > 0} {2m \over \lambda^2 + m^2}
\label{eqn:Qlatt}
\end{equation}
\begin{equation}
Q_5^{latt}=\sum_{\lambda > 0} {2m <\psi_{\lambda}|\Gamma_5|\psi_{\lambda}> 
\over \lambda^2 + m^2}
\label{eqn:Q5latt}
\end{equation}
where we have used the symmetries of the staggered action and $\Gamma_5$
is the (four-link) lattice $\gamma_5$ operator. It is then clear that in
the current lattice formulation, disconnected susceptibilities will 
{\it vanish} in the chiral limit and be proportional to $m^2$ for small
$m$. This is a direct consequence of the zero mode shift lattice
artefact \cite{vink}.  Incidently, we believe that it is this artefact which 
makes an extrapolation from lattice data extremely difficult \cite{milc,christ}. 
In view of this situation, it is important that we study how much of the
continuum behavior is already visible in the lattice data. Clearly, a 
corner stone is the Atiyah-Singer index theorem, namely the relation 
between topology and the existence of chiral fermionic zero-modes. 
Formula (\ref{eqn:Q5}) in particular comes about because in the continuum 
$<\psi_{\lambda}|\gamma_5|\psi_{\lambda}>$ is either $\pm 1$ if $\lambda
=0$ or $0$ otherwise. The lattice expression (\ref{eqn:Q5latt}) will 
closely match the continuum if we see on the lattice a clear correlation 
between small eigenvalues and large 
$r_{\lambda} \equiv <\psi_{\lambda}|\gamma_5|\psi_{\lambda}>$ 
(i.e. if we can identify chiral modes). These correlations can be
studied by drawing a plot of $r_{\lambda}$ versus $\lambda$ for all 
(low lying) eigenvalues associated with our configuration sample (as 
was done by Hands and Teper in a zero temperature SU(2) Yang-Mills theory
\cite{hands}). Here we present two examples of such plots. Fig. \ref{fig:rl6p2} 
was obtained on our sample of quenched configurations at $\beta=6.2$, 
while fig. \ref{fig:rl5p55} presents the same results in the case of 2 
flavor QCD at $\beta=5.55$. In each case, we have tried to identify the 
topology of the gauge field configurations by cooling and have used different
symbols to represent eigenvalues obtained on configurations with
different $Q_{cool}$ (see insert in fig. \ref{fig:rl6p2} and 
\ref{fig:rl5p55}). The correlation 
between large $r_{\lambda}$ and small $\lambda$ is visible on 
fig. \ref{fig:rl5p55} and very clear on fig. \ref{fig:rl6p2}.  
As expected, our computations at other values
of $\beta$ indicate that both for quenched and full QCD the correlations
deteriorate as one moves towards stronger coupling. Note that, the
relatively low value of $r_{\lambda}$ ( $\sim 0.20$ in fig. \ref{fig:rl6p2} 
and $< 0.10$ in fig. \ref{fig:rl5p55}) 
compared with the continuum value of $1.0$ follows from 
large renormalisation of the pseudoscalar operator (This is to be
expected since $\Gamma_5$, being a 4-link operator, picks up a large 
correction factor even in the mean-field approximation). 
With the knowledge of the eigenvalues and of $r_{\lambda}$, we can 
compute the disconnected susceptibilities from the formulae 
(\ref{eqn:C1d}-\ref{eqn:C5d}) and (\ref{eqn:Qlatt}-\ref{eqn:Q5latt}) 
(in our case truncated to the lowest 8 modes). The susceptibilities 
obtained in this way are compared in fig. \ref{fig:pszmdis} 
and \ref{fig:sczmdis} with those computed
with a noisy estimator in section IV. In the case of the pseudoscalar 
disconnected susceptibility (fig. \ref{fig:pszmdis}), excellent agreement 
is obtained at $\beta=5.55$ and reasonable agreement in the rest of the 
symmetric phase. A similar situation is obtained for the scalar 
susceptibilities (fig. \ref{fig:sczmdis}). 
The agreement is not expected to be as good there however 
since, even in the continuum, the explicit symmetry breaking
($ma=0.00625$ in our case) implies a sensitivity to higher eigenvalues 
(see (\ref{eqn:Q})) and dominance by the low lying modes is only expected 
to be recovered very close to the chiral limit.

Overall, we conclude that in spite of lattice artefacts, the
ingredients for a breaking of $U_A(1)$ symmetry according to the 
scenario described in section II are present in our simulations at 
finite quark mass. In particular, we have shown evidence for the
existence of topological fermionic zero-modes and have shown that 
these low lying modes are already accounting for a large part of the 
disconnected susceptibilities. In the next section, we will complete 
the argument by showing that the $U_A(1)$ symmetry breaking indeed
survives in the chiral limit once the zero-mode shift lattice artefact
is corrected for.

\begin{figure}[htb]
\epsfxsize=4in
\centerline{\epsffile{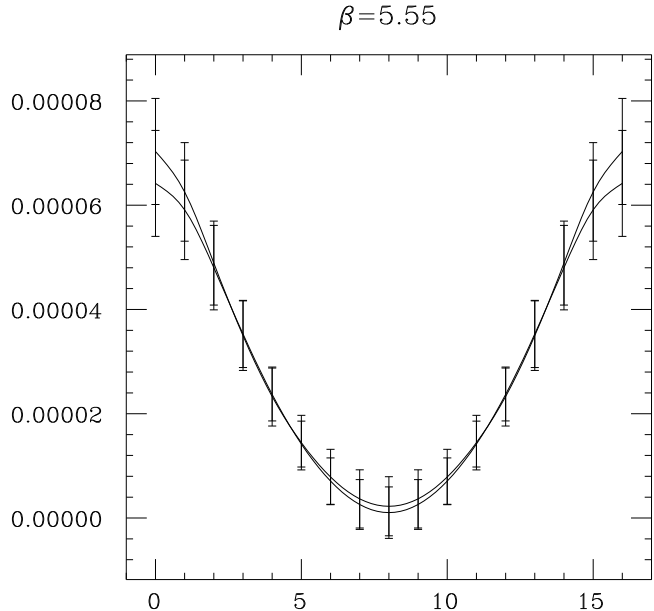}}
\caption{Comparison of the disconnected pseudoscalar correlators 
obtained from a noisy estimator (top curve) and from the truncated 
spectral decomposition of the quark propagator (bottom curve) 
at $\beta=5.55$ ($N_f=2$,$ma=0.00625$).
\label{fig:ps5p55}}
\end{figure}

Finally, it is interesting to note that not only the susceptibilities 
but the entire disconnected correlators themselves are dominated by the
low lying modes. In fig.~\ref{fig:ps5p55} ,we compare the pseudoscalar 
disconnected 
correlator at $\beta=5.55$ obtained in this way with the one computed 
with a noisy estimator. Again excellent agreement is found.
Certainly, it would be interesting to study the nature of those 
eigenmodes in greater details and for example their properties of 
localization (around instantons?).

\section{Chiral limit}

\begin{figure}[htb]
\epsfxsize=6.2in
\centerline{\epsffile{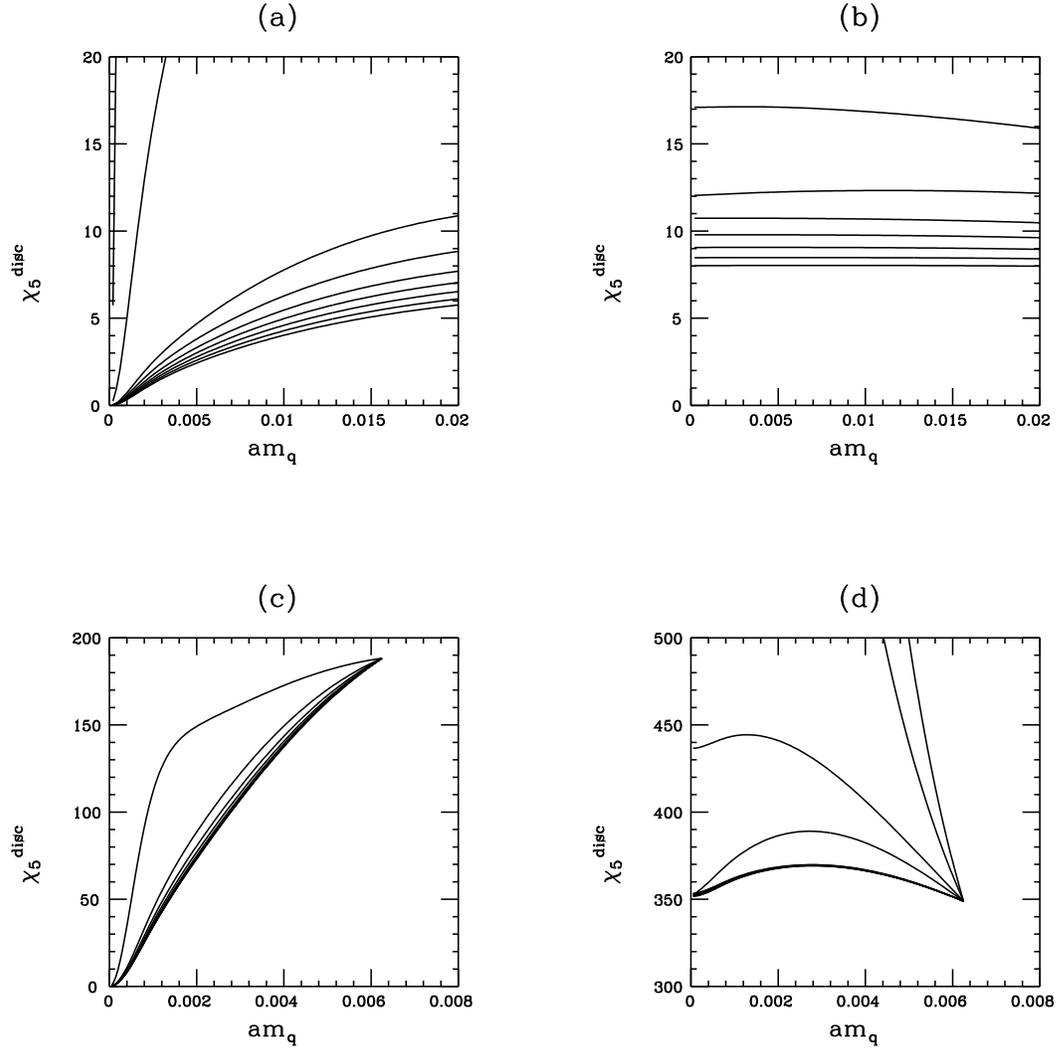}}
\vspace{2cm}
\caption{Pseudoscalar disconnected susceptibility as a function of quark mass:
\textbf{(a)} At $\beta=6.2$, in a partially unquenched approach with 0 to 8 modes 
included in the fermionic determinant (from top to bottom).  
\textbf{(b)} Same as (a) but with the Atiyah-Singer index theorem
enforced ``by hand''.
\textbf{(c)} At $\beta=5.55$, using a reweighting by partial determinants to compute the mass 
dependance, with 0 to 8 modes included (from top to bottom).
\textbf{(d)} Same as (c) but with the Atiyah-Singer index theorem
enforced ``by hand''.\label{fig:pseudo}}
\end{figure} 

In section V, we showed that the value of $\chi_5^{dis}$ at $ma=0.00625$ 
and $\beta=5.55$ could be entirely understood from the knowledge of the 
lowest 8 eigenvalues/eigenvectors of the Dirac operator on each gauge 
configuration of our sample. Since the saturation of this quantity by 
low lying modes will be even better at lower values of the quark mass 
(as can be seen from (\ref{eqn:Q5latt})), this result can be used to 
investigate the chiral limit.

\begin{figure}[htb]
\epsfxsize=6.2in
\centerline{\epsffile{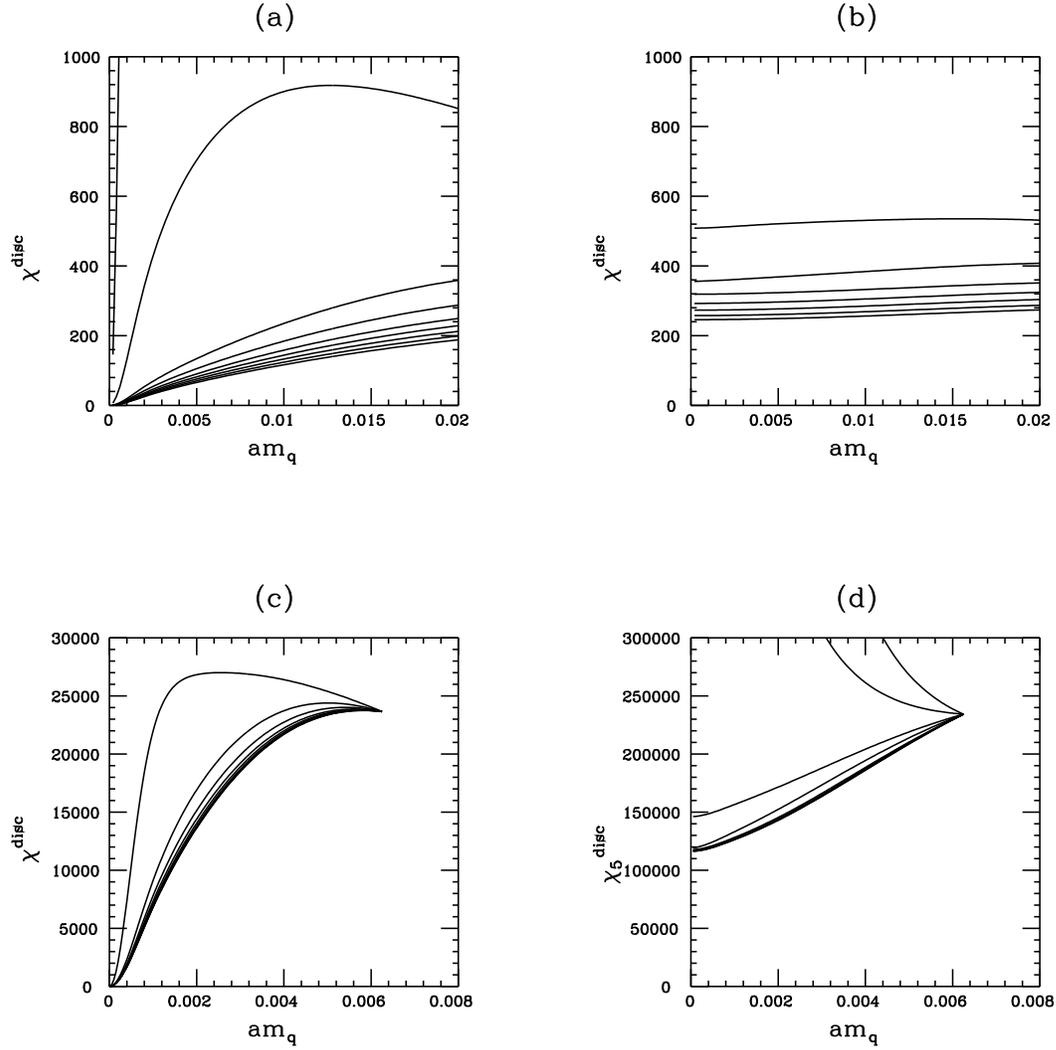}}
\vspace{2cm}
\caption{Scalar disconnected susceptibility as a function of quark mass:
\textbf{(a)} At $\beta=6.2$, in a partially unquenched approach with 0 to 8 modes 
included in the fermionic determinant (from top to bottom). 
\textbf{(b)} Same as (a) but with the Atiyah-Singer index theorem
enforced ``by hand''.
\textbf{(c)} At $\beta=5.55$, using a reweighting by partial determinants to 
compute the mass dependance, with 0 to 8 modes included (from top to bottom).
\textbf{(d)} Same as (c) but with the Atiyah-Singer index theorem
enforced ``by hand''.\label{fig:scalar}}
\end{figure}

At first, we consider a ``partially quenched approach'' where the value of 
the quark mass in the fermionic determinant is kept fixed at a value which 
we shall call $m_0$ (in our case $am_0 = 0.00625$) while a varying mass $m$ is 
introduced in the quantity we measure. A similar approach was taken in 
ref.~\cite{christ}. Since low mode dominance was obtained at $am_0 = 0.00625$, 
$\chi_5^{dis}$ can be computed from (\ref{eqn:Q5latt}) for all $ m \leq m_0 $. 
The $m$ dependance of $\chi_5^{dis}$ obtained in this way is presented in 
fig. \ref{fig:pseudo}.c
(top curve). The result depends sensitively on $m$ and vanishes at $m=0$, 
as expected from the zero-mode shift phenomenon (If there were exact zero-modes,the result would diverge like $1/m^2$). It is also easy to show that 
$\chi_5^{dis}$ will remain zero at $m=0$ in the unquenched case. For this 
purpose, we introduce partial determinants:
\begin{equation}
\Delta_k(m)= [ \prod_{n=1}^k (\lambda_n^2 + m^2) ]^{N_f/4}
\label{eqn:Dk}
\end{equation}
Successive approximations to the full QCD result are obtained by introducing 
the reweighting factor $[ \Delta_k(m) / \Delta_k(m_0) ]$ in our measurements 
(The exact answer is then obtained for $k \rightarrow k_{max}$). Fig. 
\ref{fig:pseudo}.c 
presents the results obtained in this way for k=0 to 8 (from top to bottom). 
The error bars are omitted for the clarity of the figure.
In fig. \ref{fig:scalar}.c, we 
present the result of applying the same procedure in the case of the 
disconnected scalar susceptibility. Since there is no complete dominance of 
$\chi^{dis}$ by the lowest 8 modes (see fig. \ref{fig:sczmdis}) the curves 
in fig.~\ref{fig:scalar}.c are only qualitative in character 
(as opposed to those of fig. \ref{fig:pseudo} which are 
quantitative). Quantitative details, however, will not be important in the 
discussion given below. What is important here is that $\chi^{dis}$ 
(like $\chi_5^{dis}$) vanishes in the chiral limit (although maybe with a 
slightly different approach to zero).

If taken at face value, the lattice measurements described in fig. 
\ref{fig:pseudo}.c and fig. \ref{fig:scalar}.c
would imply that the $U_A(1)$ symmetry is restored at $\beta=5.55$. However, 
we will argue that this result is the consequence of a lattice artefact and 
therefore misleading. In particular, we show below that the vanishing of 
the disconnected susceptibilities in the chiral limit is a result of the zero 
mode shift phenomenon. To do this, we turn to our quenched measurements at 
$\beta=6.2$ where the smoothness of the gauge field allows us to make the 
argument even clearer. In particular, the topology of a gauge field 
configuration can be almost unambiguously determined in this case. 
Since we are now starting from quenched configurations, the reweighting 
occurs through the partial determinants $\Delta_k$ and a ``k modes'' 
approximation to the average of an operator $A$ is given by:
\begin{equation}
\langle A \rangle_k = \int {D}U e^{-S_G} \Delta_k(m) A / Z_k
\label{eqn:Ak}
\end{equation}
\begin{equation}
Z_k = \int {D}U e^{-S_G} \Delta_k(m)
\label{eqn:Zk}
\end{equation}
The results obtained for the pseudoscalar and scalar disconnected 
susceptibilities as functions of quark mass are represented in figs. 
\ref{fig:pseudo}.a and \ref{fig:scalar}.a respectively. 
Again the disconnected susceptibilities vanish in the 
chiral limit (for any k). However, since topology can be easily identified 
in this case, it is possible to attempt to correct for the zero mode shift 
lattice artefact. We will in fact enforce the Atiyah-Singer index theorem 
``by hand'' by replacing the first $2 Q_{top}$ eigenvalues by $\lambda = 0$
\footnote{This procedure is somewhat analogous to the shifting of real 
eigenvalues for Wilson fermions \cite{bardeen}.}. After this correction is 
applied, the disconnected susceptibilities become very smooth functions of 
the quark mass, which extrapolate to non-zero values 
(figs. \ref{fig:pseudo}.b and \ref{fig:scalar}.b). 
Of course, the current approach is still partially quenched (with only up 
to 8 fermionic modes included in fig. \ref{fig:pseudo}.b and 
\ref{fig:scalar}.b) while the inclusion of the 
full fermionic determinant might make the disconnected susceptibilities 
(vanishingly) small at $\beta = 6.2$. However, the point that we want to make 
here is about the smoothness of the chiral limit, i.e. correcting for the 
zero-mode shift lattice artefact has allowed us to get around the otherwise 
apparently unavoidable consequence of a vanishing chiral limit (see fig. 
\ref{fig:pseudo}.a and \ref{fig:scalar}.a). 
A similar discussion can be given for the case of full QCD at 
$\beta=5.55$, the only difference is that the determination of the topological 
charge of gauge field configurations is now more difficult because of the 
lower value of $\beta$. Qualitatively however, the results 
(\ref{fig:pseudo}.d and \ref{fig:scalar}.d) 
are the same as shown above. Although the chiral limits obtained after 
``correction'' are only rough estimates (see fig. \ref{fig:pseudo}.d
and \ref{fig:scalar}.d), 
it is clear that the disconnected susceptibilities 
at $m=0$ are different from 0 (once the Atiyah-Singer index theorem is 
properly taken into account). In other words, the $U_A(1)$ axial symmetry 
is not restored even at $\beta=5.55$ (which corresponds to a temperature 
well above the critical temperature of the $SU(2)_L \times SU(2)_R$ 
symmetry restoration) ! The question of how large exactly the $U_A(1)$ 
symmetry breaking is (as a function of $T$ at $m=0$), however, is beyond the 
scope of this paper. The quantitative determination of the size of the 
symmetry breaking can only be addressed by using much larger lattices and 
weaker coupling or through the use of improved actions 
which better satisfy the Atiyah-Singer index theorem. Many recently introduced 
ideas need to be tested in this respect.  In this paper, we only take a modest 
first step by analyzing the case of the ``fat link'' improved quark action 
\cite{fatlink} in section VII. Other formulations which are currently under 
test include the ``perfect action'' approach \cite{hasenfratz} and the domain 
wall fermion formulation (DWF) \cite{shamir}. Some encouraging results were 
obtained recently for DWF in the context of QED in 2 dimensions \cite{vranas}.
It is also worth mentioning that dynamical Wilson fermions would not suffer 
from the ``vanishing problem'' in the chiral limit. There the zero-modes are 
shifted along the real axis and multiplication by the fermionic determinant 
ensures a smooth behavior.\footnote{In quenched QCD at zero temperature with 
Wilson fermions clear evidence for a relation between topology, real 
eigenmodes and disconnected scalar and pseudoscalar correlators was 
presented in \cite{itoh}.}
The role of DWF in this context is then to give 
a ``global'' (i.e. valid on all configurations at the same time) definition 
of the quark mass (and in particular of the point $m_q=0$).
Let us also mention that improved gauge actions would also help in so far as 
they sample configurations with smoother short distance behavior 
on which the Atiyah-Singer index theorem is better satisfied.

Finally, it is important to remember that in this paper, the issue of the 
$U_A(1)$ symmetry restoration is studied only at a single value of the 
spatial volume, namely $V=(2/T)^3$. The actual volume dependance of our 
results remains to be investigated. 

\section{Improvement of the staggered fermionic action}  

\begin{figure}[htb]
\epsfxsize=4in
\centerline{\epsffile{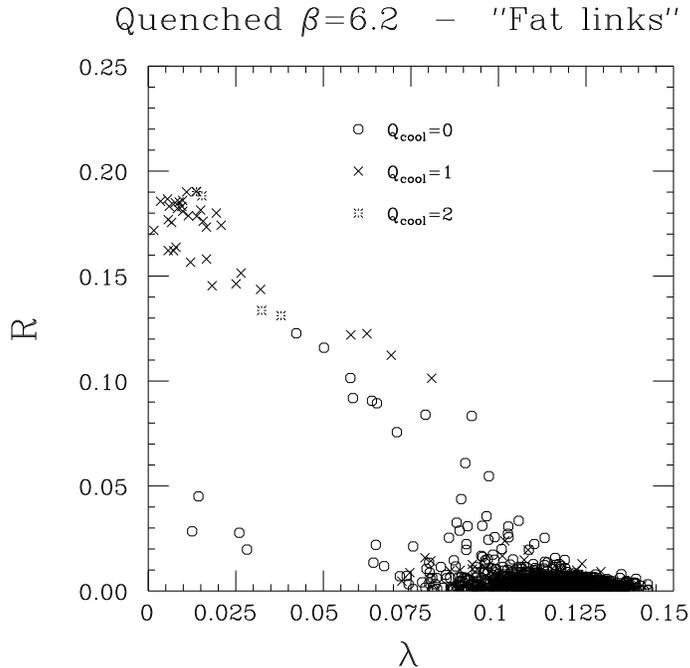}}
\caption{Eigenmode chirality versus $\lambda$ using a ``fat-link''
improved quark action on our quenched $\beta=6.2$ gauge field
configurations.\label{fig:rl6p2_stap}}
\end{figure} 

All of the results presented above clearly indicate the importance of finding 
improved quark actions which better satisfy the Atiyah-Singer index theorem. 
As already mentioned, there are many different methods which need to be tested.
These range from the traditional improvement of the staggered \cite{naik} 
and Wilson \cite{sw} fermionic action to the study of perfect actions \cite{hasenfratz} 
and the domain wall fermion formulation \cite{shamir,index}. 
That is obviously a vast program and in this paper we will limit our attention 
to a discussion of the simplest modification of the staggered action used 
above. The idea is to investigate whether the methods proposed recently to 
improve the flavor symmetry properties of the staggered action 
\cite{fatlink,sinclair} 
also help in reducing the shift of the topological zero-modes. A priori, 
one would expect that the two phenomena are related. Indeed, the continuum 
limit of the staggered theory describes 4 quark flavors, which implies a 
4-fold degeneracy of the eigenvalue spectrum in this limit. As one moves 
towards stronger coupling, the degeneracy is lifted and flavor symmetry 
breaking follows. Similarly, if there are zero modes in the continuum, 
these will be split too on the lattice.\footnote{In the case of zero-modes, 
the properties of the staggered operator imply that the splitting will be 
symmetrical. Two eigenvalues will acquire a positive imaginary part and 
the other two will be their opposites.} Therefore, in both cases improvement 
should be obtained by reducing the amount by which the eigenvalues are 
scattered. Recently, it was shown through measurements of 
$m^2_{\pi_2}-m^2_{\pi}$ that ``fat-link'' actions reduce the flavor symmetry 
breaking \cite{fatlink,sinclair}. Certainly it would be interesting to see how these actions 
modify the eigenvalue spectrum. In fig. \ref{fig:rl6p2} and
\ref{fig:rl6p2_stap}, we compare the eigenvalue 
spectrum and pseudoscalar residue $r_{\lambda} \equiv \langle \psi_{\lambda} 
| \Gamma_5 | \psi_{\lambda} \rangle$ for an ensemble of quenched configurations 
at $\beta = 6.2$ on a $(16)^3 \times 8$ lattice. 
Figure \ref{fig:rl6p2} corresponds to the usual staggered action, 
figure \ref{fig:rl6p2_stap} corresponds to a ``link+staples'' model 
where each staple carries a relative weight of $1/2$ with respect to the link.
(This action belongs to the category considered in \cite{fatlink}).
 We chose a relatively high value of $\beta$ (in the symmetric high 
temperature phase) so that topology could be easily identified. 
The symbols used in the plot 
correspond to the topological charge of each configuration (as determined by 
cooling). 8 modes have been computed per configuration.
The most noticeable feature in these two plots is associated with the cluster 
of 4 eigenvalues which is seen close to the origin in 
fig. \ref{fig:rl6p2_stap}. Those 
eigenvalues come from a single configuration and should be interpreted as 
representing a mode with low eigenvalue but zero chirality (as expected on a 
configuration with zero topological charge). In the continuum, those 4 
eigenvalues would be degenerate and $r_{\lambda}$ would be 0. With the 
standard staggered action (fig. \ref{fig:rl6p2}) these 4 eigenvalues are
much more 
dispersed and some even pick up a relatively large value of $r_{\lambda}$. 
So in this case, the improved action is clearly doing it's job: it 
significantly reduces the flavor symmetry breaking. It is also interesting 
to note that the eigenvalues just discussed are of the type that would lead 
to chiral symmetry breaking once they condense \footnote{Here we only find
one such eigenvalue since at $\beta = 6.2$, we are still 
well above $T_c$.}. Typically they are of order 
$1/V$, and since they are small but not exactly zero, they are also 
non-chiral (i.e. $r_{\lambda} = 0$). The other type of modes that we want 
to discuss are the chiral modes. In figures \ref{fig:rl6p2} and 
\ref{fig:rl6p2_stap}, these are the 
modes with $r_{\lambda}$ of the order of 0.15 or larger. These are associated 
with configurations with non-trivial topology and in the continuum would have 
$\lambda=0$. In order to preserve the distinction between the two types of 
modes and to ensure better properties of the chiral limit what we would 
at least require is that on a lattice (of finite volume) the eigenvalues 
of the chiral modes remain smaller than those of the non-chiral modes 
(which can be $O(1/V)$ when the chiral symmetry is broken). As can be seen 
on fig. \ref{fig:rl6p2_stap}, this is not yet realized 
at this stage. In other words, the ``fat link'' action doesn't necessarily 
bring the chiral modes much closer to $\lambda=0$ compared to what it does 
on other modes. The situation can be summarized by saying that this type of 
improvement only corrects the largest flavor symmetry violation. It brings 
together eigenvalues which were widely separated before but does little on 
the others. In fact, the separation that remains between the four low modes 
on figure \ref{fig:rl6p2_stap} is of the same order as the zero mode shift 
of chiral modes. 
Correcting those two effects could therefore only be achieved at a higher 
level of improvement. At the same time, other methods of improvements such 
as domain wall fermions and perfect actions should also be considered. 
It is quite conceivable that a combination of various methods 
may be necessary in the end. 

\section{Summary and conclusions} 

We have shown that important insight can be gained about the flavor
singlet dynamics of finite temperature QCD by computing the low lying 
modes of the Dirac operator. This follows from the simple result, 
derived in section II, that in the high temperature symmetric phase 
at finite volume, the scalar and pseudoscalar disconnected correlators
are entirely accounted for by the contributions of fermionic zero-modes.
Therefore, $U_A(1)$ symmetry breaking (i.e. the non-vanishing of the 
disconnected correlator) is directly linked through the Atiyah-Singer 
index theorem, to contributions from the sector of topological charge 
one in the functional integral.

This simple observation also warns us of possible difficulties with the
lattice approach, since topological properties are often not very well 
reproduced in this context. In the case of a staggered quark action, used
in this paper, the zero-mode shift lattice artefact implies that great
care has to be taken in examining the chiral limit. That there are 
difficulties in extrapolating to $m_q=0$ was already known from the 
works of refs. \cite{milc,christ} where it was shown that it is extremely 
difficult to decide between linear and quadratic fits to the data. Here 
we have gone one step further and have shown that the disconnected
susceptibilities must vanish in the chiral limit (possibly with a 
rather complicated approach) as a consequence of the zero-mode shift 
phenomenon.

Our computations also forced us to recognize this apparent restoration
of $U_A(1)$ as a lattice artefact. In section V, for example, we have
seen that our simulations indeed contain configurations with non-trivial
topological charge and that associated with these are eigenstates with
relatively large chirality and small eigenvalues. The main source of 
difficulties is that these eigenvalues are just small and not
exactly zero as would be required by the Atiyah-Singer theorem, nor are 
they so small that they can be unambiguously separated from the other  
eigenvalues which would not vanish in the continuum limit.
In fact, when we imposed the index theorem ``by hand'' and forced those
eigenvalues to vanish, we found a smooth chiral limit and disconnected
susceptibilities different from zero in the chiral limit 
(see section VI). This leads us
to the conclusion that the $U_A(1)$ symmetry is not restored at $T_c$ 
but only at a somewhat higher (possibly infinite) temperature. 
At this point however this is
still a qualitative conclusion. Quantitative questions such as 
the issue of exactly how large the symmetry breaking is as a
function of temperature will only be addressable in the context of 
improved actions or very large lattices. There is therefore an urgent 
need for studies of lattice fermion formulations which better satisfy 
the Atiyah-Singer index theorem.

In this paper, we have taken a modest first step towards investigating
improved actions by looking at the case of the ``fat-link'' formulation
\cite{fatlink}. This technique has been shown to lessen the flavor symmetry 
breaking in the mesonic spectrum. For our purposes however, we have seen
in section VII, that the improvement that it provides at the level of
the eigenvalue spectrum is too small to make a big difference in the 
identification of topological zero-modes.

Beyond this, there is much more that can be done from the knowledge of
the low lying eigenvalues and eigenvectors computed in this paper. The
spectrum itself and the distribution of eigenvalues could be compared 
with random matrix models \cite{prog}.
Since we also have the eigenvectors, their properties of localization 
possibly around instantons or other objects could be investigated as
well. In the context of finite temperature, many interesting questions
related to the change of properties of the instanton medium, the possible 
existence of instanton + anti-instanton molecules \cite{mol}, and their 
relation to quark probes deserve to be studied.

When some of the issues discussed above are settled, it will also be 
quite important to study several physical volumes rather than just 
$V=(2/T)^3$ as was done here. The fact that we have identified
topological effects and a $U_A(1)$ symmetry breaking in a relatively 
small volume doesn't necessarily mean that this will survive in the 
infinite volume limit: The nature of topological fluctuations might
depend significantly on the volume.

\section{ACKNOWLEDGEMENTS}

This work was
supported by the U.~S. Department of Energy under contract W-31-109-ENG-38,
and the National Science Foundation under grant NSF-PHY92-00148. 
The computations were performed on the CRAY C-90 at NERSC. 
We would like to thank Carleton DeTar and Edwin Laermann for informative 
conversations and the HTMCGC collaboration for the use of their configurations.

\end{document}